\documentclass[11pt]{elsarticle}
\usepackage[active]{srcltx}
\usepackage{latexsym}
\usepackage[dvips]{color}
\usepackage[iso-8859-7]{inputenc}
\usepackage{alltt}
\usepackage{graphicx}
\usepackage{amsfonts}
\usepackage{amsmath}
\usepackage{amssymb}
\usepackage{natbib}
\usepackage{psfrag}
\usepackage{indentfirst}
\usepackage{textcomp,marvosym,pifont,chemarrow}
\usepackage{layout}
\usepackage{epstopdf} 
\usepackage{float}
\usepackage{array}
\graphicspath{ {figures/} }
\usepackage[dvipsnames]{xcolor}
\usepackage{pgf,tikz}
\usepackage{mathrsfs}
\usetikzlibrary{arrows}
\usepackage{lineno}
\usepackage{hyperref}
\hypersetup{unicode,colorlinks=true,linkcolor=blue}
\newtheorem{theorem}{Theorem}[section]
\newtheorem{lemma}[theorem]{Lemma}
\newtheorem{proposition}[theorem]{Proposition}
\newproof{proof}{Proof}
\newtheorem{definition}[theorem]{Definition}
\newdefinition{rmks}{Remarks}
\newdefinition{rmk}{Remark}

\newcommand{\beq}{\begin{equation}}
\newcommand{\feq}[1]{\label{#1} \end{equation}}
\newcommand{\beqr}{\begin{eqnarray}}
\newcommand{\feqr}{\end{eqnarray}}
\def\non{\nonumber}

\newcommand{\mc}{\mathcal}

\newcommand{\rf}[1]{(\ref{#1})}
\makeatletter
\def\ps@pprintTitle{%
 \let\@oddhead\@empty
 \let\@evenhead\@empty
 \def\@oddfoot{}%
 \let\@evenfoot\@oddfoot}
\makeatother
\begin{document}
\begin{frontmatter}

\title{The short-time limit of the Dirichlet partition function and the image method}
\author[rvt]{Agapitos Hatzinikitas}
\ead{ahatz@aegean.gr}
\author[rvt]{Nikolaos Lymouris}
\ead{math14159@math.aegean.gr}
\address[rvt]{University of Aegean, School of Sciences, Department of Mathematics, Karlovasi 83200, Samos, Greece}


\begin{abstract}
In this paper we rigorously derive the $t\rightarrow 0^+$ asymptotics of the free partition function $Z_{\Omega}(t)$ for a diffusion process on tessellations of the d-dimensional Euclidean space $\mathbb{E}^d, \, d=1,2,3$ with an absorbing boundary. Utilising the path integral approach and the method of images for domains which are compatible with finite reflection subgroups of the orthogonal group $\mathbb{O}_d$, we solve this problem following a group theoretic method which was lacking from the literature. 
\end{abstract}
\end{frontmatter}
\definecolor{qqttzz}{rgb}{0.,0.2,0.6}
\definecolor{cqcqcq}{rgb}{0.7529411764705882,0.7529411764705882,0.7529411764705882}
\section{Introduction}
\label{sec0}

Pleijel \cite{P}, in 1954, established the formula 
\beqr
Z_{\Omega}(t) \stackrel{t\rightarrow 0^+}{\sim} \frac{|\Omega|}{4\pi t}-\frac{|\partial \Omega|}{8\sqrt{\pi t}}+\frac{1}{6}
\label{sec0 : eq1}
\feqr
for simply connected domains with surface area $|\Omega|$ and length of the perimeter $|\partial \Omega|$ in $\mathbb{E}^2$. Later Kac \cite{K} using a combination of probability techniques with heat equation methods justified the first two terms of \rf{sec0 : eq1} for convex domains and obtained the third term as a limiting case for convex polygonal domains. McKean and Singer \cite{MS} generalised \rf{sec0 : eq1} for smooth compact $d$-dimensional Riemannian manifolds $(\Omega,g)$ with or without $(d-1)$-dimensional Lebesgue measure of its boundary $\partial \Omega$. They proved for Dirichlet boundary conditions that the first three coefficients $c_0,\, c_1,\, c_2$ in the expansion
\beqr
Z_{\Omega}(t)=\frac{1}{(4\pi t)^{\frac{d}{2}}} \sum_{m=0}^l c_m t^{\frac{m}{2}}+o(t^{\frac{l+1}{2}})
\label{sec0 : eq2}
\feqr 
are given by
\beqr
c_0=|\Omega|, \quad c_1=-\frac{\sqrt{\pi}}{2}\int_{\partial \Omega}ds=-\frac{\sqrt{\pi}}{2} |\partial \Omega|, \quad c_2=\frac{1}{3} \int_{\Omega} K-\frac{1}{6}\int_{\partial \Omega} J
\label{sec0 : eq3}
\feqr
where $K$ is the scalar curvature (the negative spur of the Ricci tensor) and $J$ the mean curvature (the trace of the second fundamental form) at a point on $\partial \Omega$. 
\par In the case of a two dimensional compact Riemannian manifold with smooth boundary and $h<\infty$ holes, using $c_2$ and the Gauss-Bonett formula, we end up with $c_2=2\pi\chi(\Omega)/6$ where $\chi(\Omega)=2(1-h)$ is the Euler characteristic. This result coincides to the Pleijel-Kac's conjecture for a plane region with smooth boundary. If the boundary is piecewise smooth with N vertices and interior angles $\varphi_i\in (0,2\pi)$ then $c_2$ receives an extra contribution 
\beqr
\frac{1}{24\pi}\sum_{i=1}^N\frac{\pi^2 -\varphi_i^2}{\varphi_i}
\label{sec0 : eq3b}
\feqr
from the vertices. This simple expression was already known to Kac \cite{K} obtained by Dan Ray (who did not publish his result) and a transparent derivation was given by \cite{BS}.      
\par In the present paper we study the short-time asymptotics of the partition function $Z_{\Omega}(t)$ defined by 
\beqr
Z_{\Omega}(t):=\textrm{Tr}_{\Omega}\left(e^{t\Delta_D}\right)=\int_{P.B.C.}\mathcal{D}x \, e^{-S[x]}=\int_{\Omega}p_t (x,x) dx
\label{sec0 : eqc}
\feqr 
where $\Delta_D$ is the Dirichlet Laplacian, the path integral is evaluated with periodic boundary conditions (P.B.C.) and the Euclidean worldline action, for a quantum particle of mass m, is given by
\beqr
S[x]=\frac{m}{2\hbar}\int_0^{\beta}\left\|\dot{x}\right\|^2 dt.
\label{sec0 : eq4}
\feqr
Utilizing the path integral approach to the heat kernel we apply the image method in $d=1,2,3$ dimensions. The image method was pointed out by Kac in his seminal paper \cite{K} but no explicit calculation has been carried out since then along a group theory direction. The power of this method is based on the connection between the heat kernel in $\Omega$ and the heat kernel in $\mathbb{E}^d$ through the action of the reflection group. In this approach the explicit knowledge of the spectrum of the Dirichlet Laplacian, which conveys all the geometrical information of the domain, is redundant. Although the main obstruction to the eigenvalue problem of the Dirichlet Laplacian seems to be the failure of separation of variables, techniques have been developed to overcome this difficulty at least to some particular cases \cite{IZ, IZL,BM}. Then, using the Euler-Maclaurin summation formula, one can recover directly the $t\rightarrow 0^+$ asymptotics of the partition function. 
\par As far as applications is concerned, the most interesting one from the Quantum Field Theory (Q.F.T.) point of view, is the computation of the one-loop effective action for a free, real scalar field living on a tessellation\footnote{A tessellation is an infinite set of polygons (or polyhedra) fitting together to cover the plane (or space) just once, so that every side (or face) of each polygon (or polyhedron) belongs also to one other polygon (or polyhedron). No two of the polygons (or polyhedra) have common interior points.} of the Euclidean space. This is represented on a flat manifold, after imposing suitable boundary conditions (Dirichlet or Neumann), by 
\beqr
\Gamma[\phi]=-\textrm{log Det}^{-\frac{1}{2}} (-\Box)=-\frac{1}{2}\int_0^{\infty}\frac{dT}{T}\textrm{Tr}_{\Omega}(e^{T\Box})=-\frac{1}{2}\int_0^{\infty}\frac{dT}{T}\int_{P.B.C.}\mathcal{D}\textbf{X} \, e^{-S[\textbf{X}]}.
\label{sec0 : eq5}
\feqr 
The Q.F.T. action is given by 
\beqr
S_{Q.F.T.}[\phi]=\frac{1}{2}\int_{\Omega}\left\|\mathbf{\nabla}\phi\right\|^2 \, d\textbf{X} \,\,\, \textrm{where} \,\,\, d\textbf{X}=\prod_{i=1}^{d}dx_i
\label{sec0 : eq6}
\feqr 
and the path integral contains the worldline action \rf{sec0 : eq4} with $\beta$ replaced by $T$. 
\par The outline of the paper is as follows. In Section 2, we review basic definitions and fundamental theorems which allow us to give a concise introduction to the subject and prepare the mathematical background needed for proving the identification of the function $p_t^{\Omega}$ to the heat kernel on $\Omega$ (Proposition~\rf{sec2 : prop2}). We also prove the invariance of the partition function under the parabolic dilations (Proposition~\rf{sec1 : prop1}) which assures the correctness of our results up to multiplication by constants. In the $t\rightarrow 0^+$ limit of the partition function we discover an incidental formula \rf{sec1 : eq10a} which realizes the topological term for polygonal boundaries in $d=2$ dimensions, in terms of the curvature at each vertex of the polygon. 
\par In Section 3, we establish the worldline approach to compute the heat kernel for a semi-bounded domain $\Omega\subset \mathbb{E}^d$. We prove that the contribution from the direct paths to the heat kernel is identical to that of the bouncing off the boundary paths, provided that the quantum fluctuations have a linear part in time (Propositions~\rf{sec3 : prop1} and \rf{sec3 : prop2}). Extending the path space into the complement of $\Omega$, using the image method, we derive the desired heat kernel on $\Omega$. The result of this section, with slight modifications, would be applied to our problem.       
\par Section 4 presents the new group theoretic method for attacking the problem. We compute $Z_{\Omega}(t)$ in the $t\rightarrow 0^+$ limit for $d=1$, using the image method under the action of the infinite dihedral group $Dih_{\infty}$. In higher than one dimension, one faces the problem of classification of all the bounded domains dictated by the image method. This issue was partly solved in \cite{JK} and a complete solution was recorded in Coxeter's works \cite{C,CM} on discrete groups (see also \cite{RK} for an updated treatment of the subject) which are related to the classification of semi-simple Lie algebras according to Cartan and Weyl. The constraint imposed by the absence of a virtual mirror between the two given ones, after successive reflections, limits the angle of the mirrors to be of the form $\pi/m, \, \, m\in \mathbb{N}/\{1\}$. Bearing this in mind, for the infinite two-dimensional wedge, we prove that only elements of the cyclic subgroup of the finite dihedral group of order 2m, $Dih_{2m}$, contribute to the topological term (Proposition~\rf{sec2 : prop1}). In the short-time limit the result is insensitive under a truncation of the wedge (see Remark 1 on page 11). The rest of the elements of $Dih_{2m}$ contribute to the area and boundary length terms, thus Kac's result for simply connected domains in $d=2$ is recovered. We also prove, for Dirichlet boundary conditions, that expression \rf{sec1 : eq13} is the actual heat kernel on $\Omega$ in the equilateral case. The nontrivial proof can be extended to more general tessellations of the Euclidean space following similar steps. Repeating the same procedure for the infinite trihedral $(2,2,r)$ case, as was done in two dimensions, we find a closed formula for the corresponding topological term (Proposition~\rf{sec2 : prop3}). Next, for three-dimensional tessellations which are compatible with the image method, their $Z_{\Omega}(t)$ in the short-time limt is determined. It is worth noting that, for tessellations, the asymptotic behaviour  of the hyperrectangle partition function  in $d$-dimensions is possibly the only concrete result we have at present \cite{AH} (setting $s=1$).   
\section{Preliminaries}
\label{sec1}

In this section, closely following \cite{Gri}, we provide the mathematical background needed to introduce the reader to the problem and address the way one has to incorporate the method of images in order to determine the trace of the Dirichlet heat kernel. 
\par Let us denote by $\mathcal{M}$ the Euclidean space equipped with the Euclidean metric $g_{\mu \nu}=\delta_{\mu \nu}, \, \mu,\nu=1,\ldots, d$. The next theorem establishes the existence of an integral kernel for the operator $\hat{P}_t$. 
\begin{theorem}
\label{sec1 : th1}
For any $x\in \mathcal{M}$ and for any $t>0$, there exists a unique $p_{t,x}\in \mathcal{L}^2(\mathcal{M})$, such that for all $f\in \mathcal{L}^2(\mathcal{M})$, 
\beqr
\hat{P}_t f(x)\equiv e^{t\Delta}f(x)=\langle p_{t,x},f\rangle_{\mathcal{L}^2(\mathcal{M})}=\int_{\mathcal{M}}p_{t,x}(y)f(y)dy.
\label{sec1 : eq1}
\feqr
More over for any relatively compact set $K\subset \mathcal{M}$ and for any $t>0$, we have
\beqr
\underset{x\in \mathcal{M}}{\textrm{sup}}\left\| p_{t,x}\right\|_{\mathcal{L}^2(\mathcal{M})}\leq C(K,g,d)(1+t^{-\sigma})
\label{sec1 : eq2}
\feqr
where $\sigma$ is the smallest integer larger than $d/4$. 
\end{theorem}
If the function $p_{t,x}$ is defined for every $y$ then it is called the heat kernel of the manifold and satisfies a number of properties which can be found in \cite{Gri}. The heat kernel apart from being the integral kernel of the heat semigroup it can also be characterized as the minimal positive fundamental solution of the heat equation. The fundamental solution to the heat equation is defined as follows. 
\begin{definition}
\label{sec1 : def1}
Any smooth function $u$ on $\mathbb{R}_+\times \mathcal{M}$ satisfying the following conditions
\beqr
\frac{\partial u}{\partial t}&=&\Delta u \quad \textrm{in}\,\, \mathbb{R}_+\times \mathcal{M}, \label{sec1 : eq3} \\
u(t,\cdot)&\xrightarrow{\mc{D'}}& \delta_y \quad \textrm{as}\,\, t\rightarrow 0
\label{sec1 : eq4}
\feqr 
is called a fundamental solution to the heat equation at the point $y$. In the definition the class of all distributions is denoted by $\mc{D'}$. 
\end{definition}
The fundamental solution is given by the well-known Gauss-Weierstrass function 
\beqr
p_{0,t} (x,y)=\frac{1}{(4\pi t)^{d/2}}e^{-\frac{\left\|x-y\right\|^2}{4 t}}, \quad t>0, \,\, x,y\in \mathbb{R}^d.
\label{sec1 : eq4a}
\feqr
In equation \rf{sec1 : eq3} the diffusion constant, with dimensions $[L^2 T^{-1}]$, is set to one, while in \rf{sec3 : eq2} is set to $\hbar/2m$. This causes an apparent discrepancy between  expressions \rf{sec1 : eq4a} and \rf{sec3 : eq20}. Nevertheless in the calculation of the partition function we adopt the first convention. The next theorem helps us to identify whether the fundamental solution is the heat kernel, by using the ``boundary condition".
\begin{theorem}
\label{sec1 : th2}
Let $u(t,x)$ be a non-negative fundamental solution to the heat equation at the point $y\in\mathcal{M}$. If $u(t,x)\rightrightarrows 0$ as $x\rightarrow \infty$ where the convergence is uniform in $t\in(0,T)$ for any $T>0$, then $u(t,x)\equiv p_t(x,y)$. 
\end{theorem}
If $\mathcal{M}=\Omega$, where $\Omega$ is a bounded and open subset of $\mathbb{R}^d$, then every compact subset of $\mathcal{M}$ is contained in  
$\Omega_{\delta}=\{x\in\Omega:d(x,\partial\Omega)\geq \delta\}$, for some $\delta>0$. In Theorem~\rf{sec1 : th2} the convergence $x\rightarrow \infty$ in $\mathcal{M}$ means $d(x,\partial\Omega)\rightarrow 0$, namely $x\rightarrow \partial \Omega$. Also the condition that $u(t,x)$ is non-negative can be replaced by the weaker one
\beqr
\underset{k\rightarrow \infty}{\lim \textrm{sup}}\, u(t_k,x_k)\geq 0
\feqr  
for every sequence $(t_k,x_k)$ such that $t_k\rightarrow 0$ and $x_k\rightarrow x \in \mathcal{M}$. This theorem will be used in subsection (4.2) for proving that the fundamental solution, in the equilateral triangle case, is actually the heat kernel on $\Omega\subset \mathcal{M}$. 
\par The partition function defined by \rf{sec0 : eqc}, is bounded
\beqr
0<Z_{\Omega}(t)\leq \frac{p_{0,t=1}(0)}{t^{d/2}} |\Omega|
\label{sec1 : eq6}
\feqr
where $p_{0,t=1}(0)=\frac{\omega_d \Gamma(d/2)}{2(2\pi)^d}$ with $\omega_d=\frac{2\pi^{d/2}}{\Gamma(d/2)}$ the surface of the unit sphere $\mathbb{S}^{d-1}$, and $|\Omega|$ is the d-dimensional Lebesgue measure of $\Omega$. $Z_{\Omega}(t)$ enjoys an invariance property which serves as a checking rule for the correctness of its constituent terms (excluding possible multiplication constants).
\begin{proposition}
\label{sec1 : prop1}
The partition function $Z_{\Omega}$ is invariant under the parabolic dilations 
\beqr
(x,t)\mapsto (\rho x, \rho^2 t), \, \rho\in \mathbb{R}^+.
\label{sec1 : eq9a}
\feqr 
\end{proposition}
\begin{proof} 
We first prove that the heat kernels on $\Omega$ and on $\Omega'=\rho \Omega$ are related by the formula 
\beqr
p_{t'} (x',y')=\rho^{-d}p_{t} (x,y).
\label{sec1 : eq9b}
\feqr
The heat operator under \rf{sec1 : eq9a} transforms homogeneously to $\rho^2(\partial_t-\Delta_x)$ and \rf{sec1 : eq9b} satisfies the heat equation. The initial boundary condition \rf{sec1 : eq4} is preserved on $\Omega'$ provided that we multiply the heat kernel on $\Omega$ by the Jacobian factor $\rho^{-d}$. The eigenstates $\psi_m(x)$ of the Dirichlet Laplacian under the unitary dilation operator
\beqr
U(\lambda)=e^{-d\lambda/2}e^{\lambda x\cdot \nabla}, \quad e^{\lambda}=\rho
\label{sec1 : eq9c}
\feqr
transform to $\psi'_m(x')=\rho^{-d/2}\psi_m(x)$ while the eigenvalues scale like $\lambda'_m=\rho^{-2}\lambda_m$. It is then evident from $Z_{\Omega}(t)=\sum_{m=1}^{\infty}\exp(-\lambda_m t)$ that the partition function is scale invariant and therefore each term of \rf{sec0 : eq2} preserves this symmetry. 
\end{proof}
\par The asymptotic behaviour of the trace of the semigroup $\{e^{t\Delta_D}\}_{t\ge 0}$ as $t\rightarrow 0^+$, when $\Omega$ is a polytope\footnote{A polyhedral set is the intersection of a finite number of closed half-spaces. Bounded polyhedra are called polytopes.} in two dimensions with $N$ vertices, is given by combining the first two terms of \rf{sec0 : eq1} with \rf{sec0 : eq3b}. The time independent term, which is of topological origin, can be written in terms of the curvature $k_{\nu_i}$ at the vertex $\nu_i$ (see Appendix A1 for the proof) of the polygon as 
\beqr
\frac{1}{24\pi}\sum_{i=1}^N\frac{\pi^2 -\varphi_i^2}{\varphi_i}=-\frac{(N-2)}{24}+\frac{\pi}{24}\sum_{i=1}^{N}\frac{1}{\pi-k_{\nu_i}}. 
\label{sec1 : eq10a}
\feqr
 In d-dimensions the general expression for a hyperrectangle is  found in \cite{AH} and is written in terms of the mth intrinsic volume and quermassintegral. The topological term in arbitrary dimension and for a general polytope cannot be determined in closed form but it is worth noting that  
\beqr
\lim_{t\rightarrow 0^+} t^{d/2}Z_{\Omega}(t)=C_1 |\Omega|
\label{sec1 : eq11}
\feqr
where $C_1=p_{0,t=1}(0)$ and \rf{sec1 : eq11} holds without assuming that $\partial \Omega$ has finite volume. \\
\begin{rmk}
In the case of a polytope, \rf{sec0 : eq2} exhibits some interesting properties. Each term of the sum contains only one geometrical quantity characterizing the domain $\Omega$ which turns out to be invariant under the isometries of the underlying Euclidean space. This quantity represents the volume of the subspace of the boundary with codimension k and is written as $|\partial^k \Omega|$. Proposition \rf{sec1 : prop1} implies that each term of the sum is scale invariant under \rf{sec1 : eq9a} and the interchange of the sign in each term of \rf{sec0 : eq2} is necessary so that $\partial^{k}\Omega$ has opposite orientation to $\partial^{k-1}\Omega$. The topological term will be treated separately in the sequel.
\end{rmk}
\par We now speculate on the structure of the heat kernel on $\Omega$. A rigorous proof will be given by Proposition~\rf{sec2 : prop2}. Let $l=0,1,2\ldots$ be an enumeration of the virtual domains generated by successive reflections of the fundamental region in the bounding hyperplanes of $\Omega$. Let also $R_l$ be a composition of reflections that maps $\Omega$ to the region $l$. We denote by $s(l)$ the length of each element $R_l$, namely the minimum number of reflections needed to construct $R_l$. Then, for Dirichlet boundary conditions, one can write the following expression for the heat kernel on $\Omega$
\beqr
p_t^\Omega (x,y)=\sum_{l=0}^{\infty}(-1)^{s(l)}p_{0,t}(x,R_l y)\quad \forall x,y\in\Omega
\label{sec1 : eq13}
\feqr
where $R_0:= id$ is the identity matrix, $R_l y\in \bar{\Omega}$ and $\bar{\Omega}$ is the closure of $\Omega$. The asymptotic behaviour of the Dirichlet partition function in the $t\rightarrow 0^+$ limit is dominated by the diagonal elements of \rf{sec1 : eq13} which receive contributions from the heat kernels of the unbounded space for the virtual source points clustering around each vertex of the polytope (see Figure~\rf{chap2 : fig4}). This observation reduces expression \rf{sec1 : eq13} down to 
\beqr
p^v_t(x,x)\stackrel{t\rightarrow 0^+}{\sim}\sum_{n=0}^{|\mathcal{G}_v|-1}(-1)^{s(n)} p_{0,t}(x,R_n x)
\label{sec1 : eq15}
\feqr 
where $|\mathcal{G}_v|$ is the order of the reflection group at vertex $v$. Thus integrating $p_t^\Omega (x,y)$, as $x\rightarrow y$, over a suitable subdomain of $\Omega$ and after taking the $t\rightarrow 0^+$ limit we produce the corresponding expressions.
\section{The path integral approach to the heat kernel for semi-bounded domains}
\label{sec3}
We consider the flat manifold $\Omega=\mathbb{R}_+\times \partial \Omega$ with local coordinates $\textbf{X}=(x_1,x_2,\ldots,x_d)$ where the boundary $\partial \Omega=\mathbb{R}^{d-1}$ is located at $x_1=0$, $x_1\in[0,\infty)$ and $x_j\in \partial \Omega, \, j=2,\ldots,d-1$. The trajectories of a free, nonrelativistic particle of mass $m$, are described by orientable and piecewise continuously differentiable mappings: $\textbf{X}: \,\, [0,\beta]\rightarrow \Omega$. The heat kernel is the solution of the Schr\"{o}dinger equation in Euclidean time satisfying a Dirac type boundary condition at $\beta=0$
\beqr
-\hbar \frac{\partial}{\partial \beta}p(\textbf{Y},\beta; \textbf{X},0)&=& -\frac{\hbar^2}{2m}\Delta_{X}\,p(\textbf{Y},\beta; \textbf{X},0), \label{sec3 : eq2} \\
p(\textbf{Y},0; \textbf{X},0)&=& \delta^{(d)}(\textbf{X}-\textbf{Y}).
\label{sec3 : eq3}
\feqr
One method to compute the heat kernel from the Schr\"{o}dinger equation is to sum over all paths $\textbf{X}(t)$ linking the initial point $\textbf{X}_{in.}:= \textbf{X}(0)$ to the final one $\textbf{X}_{fin.}\equiv \textbf{Y}:= \textbf{X}(\beta)$, in time $\beta$
\beqr
p(\textbf{X}_{fin.},\beta;\textbf{X}_{in.},0)=\int_{\textbf{X}_{in.}}^{\textbf{X}_{fin.}}\mathcal{D}\textbf{X} \, e^{-S[\textbf{X}]}.
\label{sec3 : eq4}
\feqr  
The reader may consult \cite{BC} for the inclusion of an interaction term. The Euclidean action $S[\textbf{X}]$ in the exponent, performing a time rescaling of \rf{sec0 : eq4} by using the transformation $\tau=t/\beta, \, \tau\in[0,1]$, becomes
\beqr
S[\textbf{X}]=\frac{m}{2\beta\hbar}\int_0^{1} \left\|\dot{\textbf{X}}\right\|^2 d\tau.
\label{sec3 : eq5}
\feqr 
There are two distinctive classes of paths denoted by $\mathcal{C}_i, \, i=1,2$. The first class $\mathcal{C}_1$ has as representative the open, simple and orientable curve $\textbf{X}_{PR}(\tau)$ which starts at the point $\textbf{X}_P=\textbf{X}(0)$ and ends at $\textbf{X}_R=\textbf{X}(1)$ (see Figure~\rf{sec3 : fig1}). The contribution to the path integral from the paths which bounce off the boundary is the same as that of $\textbf{X}_{PR}(\tau)$ (see Proposition~\rf{sec3 : prop1}). The second class $\mathcal{C}_2$ has as representative the open, simple and orientable curve $\tilde{\textbf{X}}_{PR'}(\tau)$ with endpoints $\textbf{X}_P$ and $\textbf{X}_{R'}=\tilde{\textbf{X}}(1)$, respectively. The reflected paths, $\textbf{X}_{BR}$, which live in $\Omega$ can be mapped onto their images (the dashed red line in the Figure~\rf{sec3 : fig1}) and again the paths $\textbf{X}_{PB}(\tau)+\tilde{\textbf{X}}_{BR'}(\tau)$ give the same contribution as that of $\tilde{\textbf{X}}_{PR'}$. 
\begin{figure}[ht]
\centering
\includegraphics[scale=0.7]{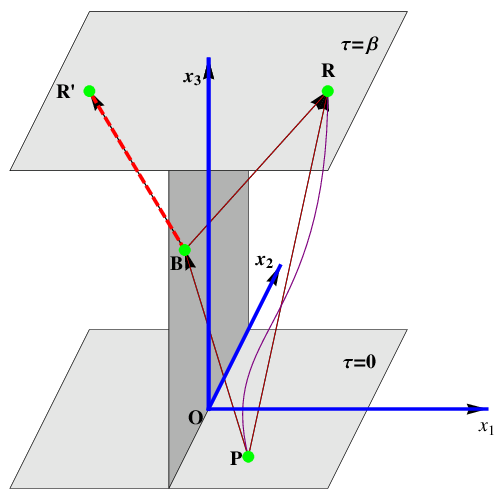} 
\caption{Solid and dashed red line segments represent the classical paths in $\Omega=\mathbb{R}_+\times \mathbb{R}^{2}$ and $\mathbb{R}^{3}$ respectively. The wiggly curve stands for the quantum trajectory in $\Omega$.} 
\label{sec3 : fig1}
\end{figure}
\par Each path can be decomposed into a fixed classical background part, satisfying the geodesic equation and subjected to appropriate homogeneous (or non-homogeneous) Dirichlet boundary conditions, plus quantum fluctuations \cite{BN}, namely,
\beqr
&& \ddot{\textbf{X}}_{cl.}(\tau)=0 
\label{sec3 : eq6}\\
&&\mathcal{C}_1: \,\, \textbf{X}(\tau)=\textbf{X}_{cl.}(\tau)+\textbf{Q}(\tau), \quad \textbf{X}(0)=\textbf{X}_P, \, \textbf{X}(1)=\textbf{X}_R,\,\textbf{Q}(0)=\textbf{Q}(1)=0 
\label{sec3 : eq7}\\
&& \ddot{\tilde{\textbf{X}}}_{cl.}(\tau)=0 
\label{sec3 : eq8}\\
&&\mathcal{C}_2: \,\, \tilde{\textbf{X}}(\tau)=\theta(\tau_1-\tau)\textbf{X}_{PB}(\tau)+\theta(\tau-\tau_1)\tilde{\textbf{X}}_{BR'}(\tau),\, 0<\tau_1<1 \, \textbf{X}_{PB}(0)=\textbf{X}_P, \, \tilde{\textbf{X}}_{BR'}(1)=\textbf{X}_{R'}, \non \\
&& \textbf{X}_{PB}(\tau_1)=\tilde{\textbf{X}}_{BR'}(\tau_1)=\textbf{X}_B, \, \textbf{Q}_{PB}(0)=\tilde{\textbf{Q}}_{BR'}(1)=0, \, \textbf{Q}_{PB}(\tau_1)=\tilde{\textbf{Q}}_{BR'}(\tau_1)=\textbf{Q}_{B}
\label{sec3 : eq9}
\feqr
where the reflection point $\textbf{X}_B$ belongs to the boundary. Solving the previous two problems we find that the paths are given by the following convex combinations of the initial and final points
\beqr
\textbf{X}(\tau)&=& (1-\tau)\textbf{X}_{P}+\tau \textbf{X}_{R}+\textbf{Q}(\tau),
\label{sec3 : eq10}\\
\tilde{\textbf{X}}_{PB}(\tau)&=& \left(1-\frac{\tau}{\tau_1}\right)\textbf{X}_P +\frac{\tau}{\tau_1}\textbf{X}_B +\textbf{Q}_{PB}(\tau),
\label{sec3 : eq11}\\
\tilde{\textbf{X}}_{BR'}(\tau)&=& \frac{(1-\tau)}{(1-\tau_1)}\textbf{X}_B +\frac{(\tau-\tau_1)}{(1-\tau_1)}\textbf{X}_{R'} +\tilde{\textbf{Q}}_{BR'}(\tau).
\label{sec3 : eq12}
\feqr
The heat kernel for the class $\mathcal{C}_1$ of paths is written as
\beqr
p_{\mathcal{C}_1,\Omega}(\textbf{X}_{R},1;\textbf{X}_{P},0)&=&e^{-\frac{m}{2\beta \hbar}\left\|\textbf{X}_R-\textbf{X}_P\right\|^2} \int_{\textbf{Q}(0)=0}^{\textbf{Q}(1)=0} \mathcal {D}\textbf{Q}\, e^{-S[\textbf{Q}]},\, S[\textbf{Q}]=\frac{m}{2\beta \hbar} \int_0^1 \left\|\dot{\textbf{Q}}(\tau)\right\|^2 d\tau
\label{sec3 : eq13}
\feqr
where
\beqr
\int_{\textbf{Q}(0)=0}^{\textbf{Q}(1)=0} \mathcal {D}\textbf{Q}\, e^{-S[\textbf{Q}]}&=& \left(\frac{m}{2\pi \beta \hbar}\right)^{d/2}.
\label{sec3 : eq14}
\feqr
\textbf{Remarks}\\
\begin{enumerate}
\item[$\bullet$] The boundary term $\langle\textbf{X}_R-\textbf{X}_P,\textbf{Q}(\tau)\rangle\bigr|_{\tau=0}^{\tau=1}$ vanishes due to the boundary conditions imposed on the fluctuations.
\item[$\bullet$] The path integral over the quantum fluctuations can be fixed by requiring that it solves the Schr\"{o}dinger equation with the Dirac boundary condition. Alternatively, one can insert a grid of $(N+1)$-ordered points over the compact time interval $[0,1]$, at equal spacing, and write the discrete quantum fluctuations in Fourier modes as in \cite{BN}
\beqr
Q^j(\tau_k)\equiv Q_k^j=\sum_{m_j=1}^{N-1}O_j^{m_j} r^j, \, O_j^{m_j}=\sqrt{\frac{2}{N}}\sin\left(\frac{km_j\pi}{N}\right), \, j=1,\ldots,d, \, k=0,\ldots, N.
\label{sec3 : eq15}
\feqr
The presence of only the sine function into the Fourier expansion \rf{sec3 : eq15} can be justified by solving the same problem for continuous time $\tau$. Due now to the orthogonality of $O_m^j$'s we can replace the path integral measure $\mathcal{D}\textbf{Q}$ by its discretized version $\prod_{k=1}^{N-1}dQ_k^j=\prod_{k=1}^{N-1}dr_k^j$ and perform the Gaussian integrals. In this way we recover the Feynman measure.   
\end{enumerate}
\begin{proposition}
\label{sec3 : prop1}
The following identity for the heat kernels holds
\beqr
p_{\mathcal{C}_1,\Omega}(\textbf{X}_{R},1;\textbf{X}_{P},0)=p_{\tilde{\mathcal{C}}_1,\Omega}(\textbf{X}_{R},1;\textbf{X}_{P},0)
\label{sec3 : eq18}
\feqr
where $\mathcal{C}_1$ consists of direct paths with endpoints $\{\textbf{X}_{P},\textbf{X}_{R}\}$ while $\tilde{\mathcal{C}}_1$ is composed by paths which bounce off the boundary but have the same endpoints as previously.
\end{proposition}
\begin{proof}
The action for $\tilde{\mathcal{C}}_1$ paths, is
\beqr
S_{\tilde{\mathcal{C}}_1}[\textbf{X}]&=&S[\textbf{X}_{PB}]+S[\textbf{X}_{BR}] \non \\
&=&\frac{m}{2\beta \hbar\tau_1} \left\|\textbf{X}_B-\textbf{X}_P+\textbf{Q}_B \right\|^2+\frac{m}{2\beta \hbar}\int_0^{\tau_1} \left\|\dot{\textbf{Q}}(\tau)\right\|^2 d\tau \non \\
&+&\frac{m}{2\beta \hbar(1-\tau_1)} \left\|\textbf{X}_R-\textbf{X}_B-\textbf{Q}_B\right\|^2+\frac{m}{2\beta \hbar}\int_{\tau_1}^1 \left\|\dot{\textbf{Q}}(\tau)\right\|^2 d\tau 
\label{sec3 : eq19}
\feqr
where the boundary terms $\langle\textbf{X}_B-\textbf{X}_P,\textbf{Q}_{PB}(0)\rangle$ and $\langle\textbf{X}_R-\textbf{X}_B,\textbf{Q}_{BR}(1)\rangle$ again vanish due to the boundary conditions of the fluctuations. 
The heat kernel is then given by
\beqr
p_{\tilde{\mathcal{C}}_1,\Omega}(\textbf{X}_{R},1;\textbf{X}_{P},0)&=& \left(\int_{\textbf{X}_P}^{\textbf{X}_B}\mathcal{D}\textbf{X}_{PB}\, e^{-S[\textbf{X}_{PB}]}\right)\left(\int_{\textbf{X}_B}^{\textbf{X}_R}\mathcal{D}\textbf{X}_{BR}\, e^{-S[\textbf{X}_{BR}]}\right)\non \\
\!\!\!\!&=&\!\!\!\!\left(\frac{m}{2\beta \hbar}\right)^d\!\!\! \left(\frac{1}{\tau_1(1-\tau_1)}\right)^{d/2}\!\!\!\!\int_{\mathbb{R}^d}\!\!\!\! e^{-\frac{m}{2\beta \hbar}\left(\frac{1}{\tau_1}\left\|\textbf{X}_B-\textbf{X}_P+\textbf{Q}_B \right\|^2+\frac{1}{(1-\tau_1)} \left\|\textbf{X}_R-\textbf{X}_B-\textbf{Q}_B\right\|^2\right)} d\textbf{Q}_B\non \\
&=& \left(\frac{m}{2\pi \beta \hbar}\right)^{d/2}e^{-\frac{m}{2\beta \hbar}\left\|\textbf{X}_R-\textbf{X}_P\right\|^2}
=p_{\mathcal{C}_1,\Omega}(\textbf{X}_{R},1;\textbf{X}_{P},0).
\label{sec3 : eq20}
\feqr
The integral in the second line can be performed by minimizing the exponential and making a displacement transformation in $Q^j_B$.
\end{proof}
\textbf{Remarks}
\begin{enumerate}
\item[$\bullet$] We used the following decompositions for the quantum fluctuations
\beqr
\textbf{Q}_{PB}(\tau)&=&\textbf{Q}(\tau)+\frac{\tau}{\tau_1} \textbf{Q}_{B}, \,\, \textbf{Q}(0)=\textbf{Q}(\tau_1)=0 \\
\textbf{Q}_{BR}(\tau)&=&\textbf{Q}(\tau)+\frac{1-\tau}{1-\tau_1} \textbf{Q}_{B}, \,\, \textbf{Q}(\tau_1)=\textbf{Q}(1)=0.
\label{sec3 : eq22}
\feqr
The verification of these expressions is explained by Proposition~\rf{sec3 : prop2}.
\item[$\bullet$] The constants are generated following the same method as that of the quantum fluctuations for the class $\mathcal{C}_1$. 
\item[$\bullet$] The integration of $\textbf{Q}_B$ over $\mathbb{R}^d$ is needed since the path integral should be independent of the location of the boundary on $x_1$-axis.
\end{enumerate}
\begin{proposition}
\label{sec3 : prop2}
The asymptotic behaviour of the paths $Q^j_{PB}(\tau)$, in the $Q^j_B\rightarrow 0$ limit, is given by
\beqr
Q^j_{PB}(\tau)\sim \sqrt{\frac{2}{\tau_1}}\sin(\lambda_{m_j} \tau)+(-1)^{m_j}\frac{\tau}{\tau_1}Q^j_B\cos(\lambda_{m_j} \tau), \, \lambda_{m_j}=\frac{\pi m_j}{\tau_1}.
\label{sec3 : eq23}
\feqr  
\end{proposition}
\begin{proof}
The path $Q^j_{PB}(\tau)$ satisfies the problem
\beqr
\ddot{Q}_{PB}^j(\tau)&=&-\lambda^2_{PB} Q^j_{PB}(\tau), \quad Q^j_{PB}(0)=0, \, Q^j_{PB}(\tau_1)= Q^j_{B}
\label{sec3 : eq24}
\feqr
with solution
\beqr
Q^j_{PB}(\tau)&=&\frac{Q^j_{B}}{\sin(\lambda_{PB}\tau_1)}\sin(\lambda_{PB}\tau).
\label{sec3 : eq25}
\feqr
For simplicity we consider the one-dimensional case. Let $\lambda_{PB}=\lambda_{m}f(Q_B)$ with $f(0)=1$ since $\lambda_{PB}\rightarrow \lambda_{m}$ in the $Q_B \rightarrow 0$ limit. Requiring
\beqr
\lim_{Q_B \rightarrow 0}\left(\frac{Q_{B}}{\sin(\lambda_{PB}\tau_1)}\right)=\sqrt{\frac{2}{\tau_1}},
\label{sec3 : eq26}
\feqr
applying L'Hopital's rule and Taylor expanding $f(Q_B)$ around the point $Q_B=0$, we find that
\beqr
\lambda_{PB}-\lambda_m=\frac{(-1)^m}{\sqrt{2\tau_1}}Q_B.
\label{sec3 : eq27}
\feqr
Taylor expanding the sines functions of the solution around $\lambda_m$, for fixed time, we obtain the desired result. A similar expression is derived for the path $Q^j_{BR}(\tau)$ with the only change that the time spacing is now $1-\tau$. The above asymptotic solution can be shown to satisfy equation \rf{sec3 : eq24}. 
\par Let us denote by $Q^j_{B}(\tau)=(-1)^{m_j}Q^j_B\cos(\lambda_{m_j} \tau)$ the quantum fluctuations related to the boundary condition $Q^j_{B}(\tau_1)=Q^j_{B}$. Expanding $Q^j_{PB,k}$ in Fourier modes we can realize that the corresponding transformation from $dQ^j_{PB,k}$ to $dr^j, ds^j$ is not orthogonal and the calculation of the path integral breaks down. To overcome this difficulty we average $Q^j_{B}(\tau)$ over the time interval $[0,\tau_1]$ and the final expression of the fluctuations is identical to \rf{sec3 : eq22}.   
\end{proof}
The heat kernel for the semi-bounded space is given by
\beqr
p_t(x,y)= p_{0,t}(\left\|\mathbf{X}-\mathbf{Y}\right\|)- p_{0,t}(\left\|\mathbf{X}+\mathbf{Y}\right\|)
\label{sec3 : eq28}
\feqr
where $p_{0,t}$ is the fundamental solution \rf{sec1 : eq4a}.
\section{The $Z_{\Omega}(t)$ in the $t\rightarrow 0^+$ limit}
\subsection{ The \texorpdfstring{$d=1$}{d1} case}
\label{sec2}

In $d=1$, the reflection group, $\mathcal{G}$, coincides to the infinite dihedral group $\textrm{Dih}_{\infty}$ with defining relations
\beqr
R_{a_1}^2=R_{a_2}^2=I
\label{sec2 : eq1}
\feqr
where $R_{a_i}$'s are involuntary transformations and represent reflections with respect to the boundary points of $\Omega\subset \mathbb{R}$. In this presentation $\textrm{Dih}_{\infty}$ is the free product of two $\mathbb{Z}/2\mathbb{Z}$ \footnote{Alternatively $\textrm{Dih}_{\infty}$ is isomorphic to the semidirect product $\mathbb{Z}\rtimes \mathbb{Z}/2\mathbb{Z}=<g,h| h^2=I, \, hgh=g^{-1}>$.}. 
Assuming that $\Omega=(0,L)$, for every $y \in \Omega$, we generate the following two infinite sequences of virtual source points depending on whether we start reflection from the left or right fixed point of the isometry $R_{a_i}$ (see Figure~\rf{chap2 : fig1}).
\begin{table}[h]
\centering
\begin{tabular}{|c|c||c|c|}\hline
\multicolumn{4}{|c|}{\bfseries The virtual image points of $y$} \\ \hline 
\multicolumn{2}{|c||}{\itshape Left fixed point} & \multicolumn{2}{c|}{\itshape Right fixed point} \\ \hline 
\itshape Group Element & \itshape Location of virtual image point & \itshape Group Element & \itshape Location of virtual image point  \\ \hline \hline
$R_{a_2}$ & -y & $R_{a_1}$ & -y+2L \\ \hline
$R_{a_1}\cdot  R_{a_2}$ & y+2L & $R_{a_2}\cdot R_{a_1}$ & y-2L \\ \hline
$R_{a_2}\cdot R_{a_1}\cdot R_{a_2}$ & -y-2L & $R_{a_1}\cdot R_{a_2}\cdot R_{a_1}$ & -y+4L \\ \hline
$(R_{a_1}\cdot R_{a_2})^2$ & y+4L & $(R_{a_2}\cdot R_{a_1})^2$ & y-4L \\ \hline
$\vdots$ & $\vdots$ & $\vdots$ & $\vdots$ \\ \hline
\end{tabular}
\caption{The locations of the virtual image points derived from the action of the elements of the $\textrm{Dih}_{\infty}$ group.}
\end{table}
\begin{figure}[h]
\centering
\includegraphics{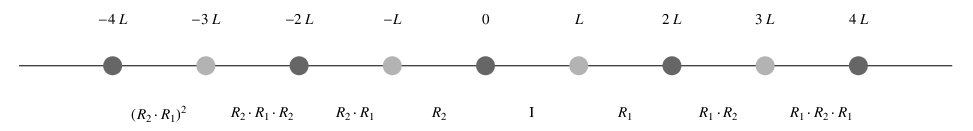} 
\caption{The fundamental region $\Omega=(0,L)$ and the virtual domains generated by the elements $R_i:= R_{a_i}$ of the infinite dihedral group.} 
\label{chap2 : fig1}
\end{figure}

The heat kernel in this case is written as 
\beqr
p_t(x,y)= p_{0,t}(\left\|x-y\right\|)-\sum_{n\in \mathbb{Z}} p_{0,t}(\left\|x+y+2nL\right\|)+\sum_{n\in \mathbb{Z}\setminus \{0\}}p_{0,t}(\left\|x-y+2nL\right\|)
\label{sec2 : eq3}
\feqr
where in the first sum we have grouped the contributions of odd number of group elements while in the second sum that of even number of group elements. Relation \rf{sec1 : eq13} then becomes
\beqr
p_t(y,y)= \frac{1}{\sqrt{4\pi t}}\left(1-\sum_{n\in \mathbb{Z}} e^{-\frac{\left\|y+nL\right\|^2}{t}}+2\sum_{n\in \mathbb{N}\setminus \{0\}} e^{-\frac{\left\|nL\right\|^2}{t}}\right).
\label{sec2 : eq4}
\feqr
\par Integrating \rf{sec2 : eq4} over the fundamental domain, taking the $t\rightarrow 0^+$ limit and maintaining $t^{-1/2}$ terms we obtain 
\beqr
Z_{\Omega}(t)&\sim& \frac{1}{\sqrt{4\pi t}}\int_0^L dy -\lim_{t\rightarrow 0^+}\left(\frac{1}{\sqrt{4\pi t}}\sum_{n\in \mathbb{Z}}\int_0^L e^{-\frac{\left\|y+nL\right\|^2}{t}} dy\right) +2\lim_{t\rightarrow 0^+}\left(\frac{1}{\sqrt{4\pi t}}\sum_{n\in \mathbb{N}\setminus \{0\}} \int_0^L e^{-\frac{\left\|nL\right\|^2}{t}} dy\right) \non \\
&=& \frac{L}{\sqrt{4\pi t}} -2\lim_{t\rightarrow 0^+}\left(\frac{1}{4} \textit{erf}\left(\frac{L}{\sqrt{t}}\right)\right)+\sum_{n\in \mathbb{N}\setminus \{0\}} \delta(n) 
=\frac{L}{\sqrt{4\pi t}}-\frac{1}{2}.
\label{sec2 : eq5a} 
\feqr
\subsection{The \texorpdfstring{$d=2$}{d2} case}

The case $d=2$ is more involved. Let $\theta$ be the angle between two intersecting mirror rays in $\mathbb{R}^2$. If we require the absence of a virtual mirror ray between the two given ones, after successive reflections of the fundamental domain, then $\theta=\pi/q$ where $q\in \mathbb{N}/ \{1\}$. This remark facilitates the enumeration of bounded tessellations of the plane through reflections. For triangular domains with angles $\pi/p, \, \pi/q, \, \pi/r$ and $p, q, r \in \mathbb{N}/ \{1\}$, we have
\beqr
\frac{\pi}{p}+\frac{\pi}{q}+\frac{\pi}{r}=\pi.
\label{sec2 : eq6}
\feqr
Relation \rf{sec2 : eq6} is satisfied for the congruent equilateral triangles $(3, 3,3)$, the isosceles right triangles $(2, 4, 4)$ and the bisected equilateral triangles $(2, 3, 6)$. Moreover, the only other admissible polygon is the rectangle.
\par We now study the case of an infinite plane wedge (see Figure~\rf{chap2 : fig2}) since this is the guiding principle for investigating the topological term. If the angle $\theta=\pi/m, \, m\in\mathbb{N} \setminus \{1\}$ then there is a unique, up to isomorphism, group generated by two involutions $R_{\alpha_1}, \, R_{\alpha_2}$ such that their product $R_{\alpha_1}\cdot R_{\alpha_2}$ has order $m$. The group is denoted by $\textrm{Dih}_{2m}$, called the dihedral group of order $2m$ and has the presentation
\beqr
\textrm{Dih}_{2m}=<R_{\alpha_1}, \, R_{\alpha_2}|\,\, R_{\alpha_1}^2=R_{\alpha_2}^2=(R_{\alpha_1}\cdot R_{\alpha_2})^m=I>.
\label{sec2 : eq7}
\feqr
An alternative way to define $\textrm{Dih}_{2m}$ would be the semidirect product
\beqr
\textrm{Dih}_{2m}=\mathbb{Z}/m\mathbb{Z}\rtimes \mathbb{Z}/2\mathbb{Z}
\label{sec2 : eq7a}
\feqr
where if  $h$ generates $\mathbb{Z}/2\mathbb{Z}$ then $hgh=g^{-1}$, $\forall g\in \mathbb{Z}/m\mathbb{Z}$. In this set up the presentation reads
\beqr
\textrm{Dih}_{2m}=<g,\, h|\,\, h^2=g^m=(hg)^2=I>
\label{sec2 : eq7b}
\feqr
and it can be proved that \rf{sec2 : eq7b} is equivalent to \rf{sec2 : eq7} under the substitutions $R_{\alpha_1}=h$ and $R_{\alpha_2}=hg$.
\begin{figure}[h]
\centering
\includegraphics[scale=0.6]{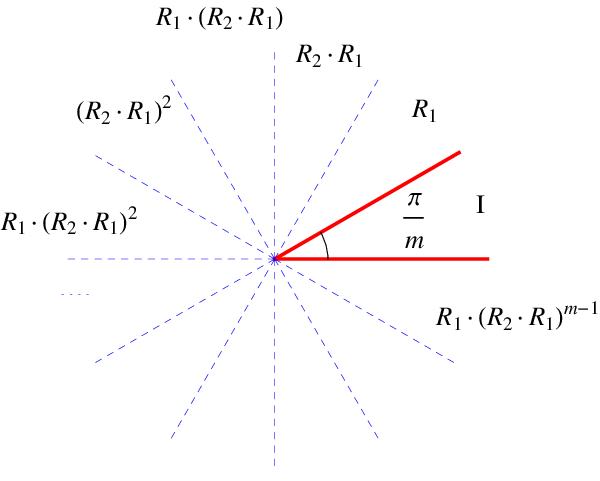} 
\caption{The fundamental region of the infinite wedge with $\theta=\pi/m$ and the virtual domains generated by the elements $R_i:= R_{a_i}$ of the group $\textrm{Dih}_{2m}$.} 
\label{chap2 : fig2}
\end{figure}
From \rf{sec2 : eq7a} observe that the dihedral group has the cyclic group $\mathcal{C}_m=\mathbb{Z}/m\mathbb{Z}$ as a subgroup. Also $\textrm{Dih}_{2m}$ is the group of automorphisms of the regular planar $m-$gon. 
\par Now we adopt the traditional approach to finite reflection groups using root systems. Let us choose the normalised root vectors to be  
\beqr
a_1 = (-\sin \theta, \cos \theta)^{\scriptscriptstyle\top}, \quad
a_2 = (0, -1)^{\scriptscriptstyle\top} 
\label{sec2 : eq8b}
\feqr
where ``$^{\scriptscriptstyle\top}$'' stands for the transpose. In general, the reflection of a vector $r\in \mathbb{E}^d$ through a fixed hyperplane $H_a=\{r\in \mathbb{E}^d\, | <r,a>=0\}$ for a given vector $a\neq 0$, is defined by
\beqr
R_{a}r=r-2\frac{<a,r>}{<a,a>}a
\label{sec2 : eq9}
\feqr
where $R_{a}r=r$, if $r\in H_a$, and $R_{r}r=-r$. The reflection matrices in $d=2$ are
\beqr
 R_{a_1}(\theta)=\left( \begin{array}{cc} \cos (2\theta) & \sin (2\theta) \\ \sin (2\theta) & -\cos (2\theta) \end{array}\right)  \quad 
\textrm{and} \quad R_{a_2}=\left(\begin{array}{cc} 1 & 0 \\ 0 & -1 \end{array}\right) .
\label{sec2 : eq10}
\feqr 
A simple computation reveals that the element $R_{a_2}\cdot R_{a_1}$, which represents a rotation through twice the angle between the two rays, generates the cyclic group $\mathcal{C}_m$ with elements
\beqr
(R_{a_2}\cdot R_{a_1})^k (\theta)=\left( \begin{array}{cc} \cos (2k\theta) & \sin (2k\theta) \\ -\sin (2k\theta) & \cos (2k\theta) \end{array}\right), \quad k=0,1,2,\ldots,m-1.
\label{sec2 : eq11}
\feqr  
\begin{proposition}
\label{sec2 : prop1}
In two dimensions, only the elements of the cyclic group $\mathcal{C}'_m \subset \textrm{Dih}_{2m} $ contribute to the topological term which is present in the short-time limit of the partition function. 
\end{proposition}
\begin{proof}
The contribution of the elements \rf{sec2 : eq11}, excluding the identity element (therefore the notation $\mathcal{C}'_m$), to the heat kernel at coincident points is
\beqr
p_t(r,r)=\frac{1}{4\pi t}\sum_{k=1}^{m-1}e^{-\frac{r^2}{t}\sin^2\left(\frac{k\pi}{m}\right)} 
\label{sec2 : eq12}
\feqr
which after integration over the infinite angular region in polar coordinates gives
\beqr
\int_{0}^{\infty}p_t(r,r)r dr\int_{0}^{\pi/m} d\theta=\frac{1}{8m}\sum_{k=1}^{m-1}\frac{1}{\sin^2\left(\frac{k\pi}{m}\right)}=\frac{1}{24m}(m^2-1). 
\label{sec2 : eq13}
\feqr
The computation of the finite series in the right-hand side of the first equality of \rf{sec2 : eq13} is given in Appendix B.
\end{proof}
\textbf{Remarks}
\begin{enumerate}
\item If we truncate the infinite wedge by using a circular-arc of radius $R$, then the topological term remains intact. The claim in this case can be proved by using the relation 
\beqr
Z_{\Omega}(t)=\sum_{l,n=1}^{\infty}e^{-ta_{l,n}^2}
\label{sec2 : eq14}
\feqr
where $a_{l,n}^2$ are the eigenvalues of the Laplacian (a suitable rescaling of the distance is taken into account) on the finite wedge when Dirichlet boundary conditions are imposed. The eigenvalues are determined by the roots of $J_{l m}(a_{l,n})=0$ where $J_{\nu}(z)$ is the Bessel function of the first kind of positive integer order $\nu=l m$. In the $t\rightarrow 0^+$ limit, \rf{sec2 : eq14} is dominated by the contribution of the large zeros, therefore we make use of the fact that $J_{\nu}(\nu \sec\beta)$ has a positive root wherever \cite{WAT}
\beqr
\nu(\tan \beta -\beta)+\chi=n\pi
\label{sec2 : eq14ab}
\feqr
where $n$ is any positive integer (unity possibly excepted), and, for large $\nu$, $\pi/6<\chi<\pi/4$. Therefore we end up with the following relation of finite differences
\beqr
\Delta n=\frac{\nu}{\pi} \sin \beta \, \Delta(\sec \beta).
\label{sec2 : eq14ac}
\feqr
Using the Euler-Maclaurin sum formula \cite{AD} and taking the $t\rightarrow 0^+$ limit at the end we recover the result
\beqr
Z_{\Omega}(t) \stackrel{t\rightarrow 0^+}{\sim} \frac{1}{8 m t}-\frac{(2m+\pi)}{8m\sqrt{\pi t}}+\frac{(m^2-1)}{24m}
\label{sec2 : eq14ad}
\feqr
\item If we approximate a circle by an inscribed regular polygon of $N$ vertices then \rf{sec1 : eq10a} becomes
\begin{align}
\frac{1}{24\pi}\sum_{i=1}^N \frac{\pi^2- \phi_i^2}{\phi_i}=\frac{1}{6} \left(\frac{N-1}{N-2}\right).
\label{sec2 : eq13a}
\end{align} 
In \rf{sec2 : eq13a} as the number of vertices tends to infinity the sum converges to the geometrical constant $1/6$ of \rf{sec0 : eq1}.
\end{enumerate}
We now study in detail the equilateral triangle $\Omega:=\triangle$ of side $L$ and height $h=L\sqrt{3}/2$. The symbol $\triangle$ should not be confused to the one used to denote the Dirichlet Laplacian. In proving that $p_t^{\triangle}$ is actually the heat kernel of the equilateral triangle we will need the following three lemmas.
\begin{lemma}
\label{sec2 : lem1a}
Let $u(t,x)$ be a non-negative smooth function on $\mathbb{R}^+\times \mathcal{M}$ satisfying $\int_{\mathcal{M}}u(t,\cdot)dx\leq 1$ and such that $u(t,\cdot)\xrightarrow{\mc{D'}} \delta_y$ as $t\rightarrow 0$. Then for every open set $\Omega\Subset \mathcal{M}$ and every continuous and bounded function $f\in C_b(\Omega)$ holds
\beqr
\int_{\Omega}u(t,x)f(x)dx\rightarrow \left\{\begin{array}{ccc} f(y), & \textrm{if} & y\in\Omega \\
0, & \textrm{if} & y\in \mathcal{M}/\bar{\Omega}\end{array}\right.
\label{sec2 : 13b}
\feqr
as $t\rightarrow 0$. In \rf{sec2 : 13b} the symbol $A\Subset B$ means ``compact inclusion", namely the closure $\bar{A}$ of the set A is compact and $\bar{A}\subset B$. 
\end{lemma}
\begin{lemma}
\label{sec2 : lem1}
Let $M$ be a Riemannian manifold, $a,b\in\mathbb{R}$. Suppose $\{u_k\}_{k=1}^\infty$ is a non-decreasing sequence of solutions of the heat equation on $(a,b)\times M$, such that  
\beqr
\int_M |u_k(t,x)|dx\leq C,
\label{sec2 : 14a}
\feqr
where the constant $C$ is independent of $k$ and $t\in(a,b)$. Then  $u=\lim_{k\rightarrow\infty}u_k$ is a smooth solution of the heat equation on $M\times (a,b)$ and $u_k\rightarrow u$  uniformly on compacta together with derivatives of all orders.
\end{lemma}
The proof can be found in \cite{Dod}.
\begin{lemma}
\label{sec2 : lem2}
For every $\varepsilon>0$ the function $\nabla p_t(x)$ is bounded on $|x|\geq\varepsilon, t>0$ and  
\beqr
\sup_{\begin{subarray}{c}
t>0\\
|x|\geq\varepsilon
\end{subarray}}|\nabla p_t(x)|\leq \frac{\textrm{const.}}{\varepsilon^{d+1}}
\label{sec2 : 14b}
\feqr
where $p_t(x)$ is the \textit{Gauss-Weierstrass} function \rf{sec1 : eq4a}.
\end{lemma}
\begin{proof}
Substituting $\tau=t/|x|^2$ in 
\beqr
|\nabla p_t(x)|=\frac{1}{(4\pi t)^{d/2}}\frac{|x|}{2t}e^{-\frac{|x|^2}{4t}}
\label{sec2 : eq13m}
\feqr
we obtain
\beqr
\sup_{t>0}|\nabla p_t(x)|=\sup_{\tau >0}\frac{1}{\left(4\pi \tau |x|^2\right)^{d/2}}\frac{1}{2\tau |x|}e^{-\frac{1}{4\tau}}=\frac{\textrm{const.}}{|x|^{d+1}}.
\label{sec2 : eq13n}
\feqr
\end{proof}
\begin{proposition}
\label{sec2 : prop2}
The function 
\beqr
p_t^\triangle (x,y)=\sum_{l=0}^{\infty}(-1)^{s(l)}p_{0,t}(x,R_l y)\quad x,y\in\triangle
\label{sec2 : eq13b}
\feqr
is the heat kernel of the equilateral triangle.
\end{proposition}
\begin{proof}
Proceeding stepwise, we shall prove:
\begin{enumerate}
\item The function $p_t^\triangle(x,y)$ is a smooth solution of the heat equation on $(0,\infty)\times \triangle$.

We first split the sum into two groups of sums carrying the same sign. Let  $p_t^{+\triangle}(x,y)=\sum_{l=0}^{\infty} p_{0,t}(x,R'_l y)$ be the sum of positive terms where $R'_0=id$. The partial sums $\sum_{l=0}^{k} p_{0,t}(x,R'_l y)$ constitute an increasing sequence of solutions to the heat equation on $\triangle\times (0,\infty)$. If we consider an increasing sequence of triangles on the plane which is similar and concentric to the initial triangle (see Figure~\rf{sec2 : fig1a}), then the plane is divided into sectors $T_n, n=0,1,2\ldots$ where  $T_0=\triangle$. Each sector consists of $18n-3$ virtual domains from which $9n-1$, at most, correspond to terms with positive sign. 
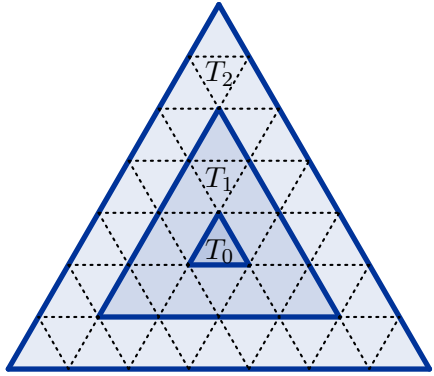
\begin{figure}[h]
\centering
\begin{tikzpicture}[scale=0.4,line cap=round,line join=round,>=triangle 45,x=1.0cm,y=1.0cm]

\clip(-3.,-4.246446646847413) rectangle (13.,9.363293836931012);
\fill[line width=2pt,color=qqttzz,fill=qqttzz,fill opacity=0.10000000149011612] (4.,0.) -- (6.,0.) -- (5.,1.7320508075688776) -- cycle;
\fill[line width=2pt,color=qqttzz,fill=qqttzz,fill opacity=0.10000000149011612] (1.,-1.7320508075688776) -- (9.,-1.7320508075688776) -- (5.,5.196152422706631) -- cycle;
\fill[line width=2pt,color=qqttzz,fill=qqttzz,fill opacity=0.10000000149011612] (-2.,-3.4641016151377553) -- (12.,-3.4641016151377553) -- (5.,8.660254037844386) -- cycle;
\draw [line width=2pt,color=qqttzz] (4.,0.)-- (6.,0.);
\draw [line width=2pt,color=qqttzz] (6.,0.)-- (5.,1.7320508075688776);
\draw [line width=2pt,color=qqttzz] (5.,1.7320508075688776)-- (4.,0.);
\draw [line width=2pt,color=qqttzz] (1.,-1.7320508075688776)-- (9.,-1.7320508075688776);
\draw [line width=2pt,color=qqttzz] (9.,-1.7320508075688776)-- (5.,5.196152422706631);
\draw [line width=2pt,color=qqttzz] (5.,5.196152422706631)-- (1.,-1.7320508075688776);
\draw [line width=2pt,color=qqttzz] (-2.,-3.4641016151377553)-- (12.,-3.4641016151377553);
\draw [line width=2pt,color=qqttzz] (12.,-3.4641016151377553)-- (5.,8.660254037844386);
\draw [line width=2pt,color=qqttzz] (5.,8.660254037844386)-- (-2.,-3.4641016151377553);
\draw [line width=0.8pt,dotted] (1.,1.7320508075688776)-- (4.,-3.4641016151377553);
\draw [line width=0.8pt,dotted] (2.,3.4641016151377553)-- (6.,-3.4641016151377553);
\draw [line width=0.8pt,dotted] (4.,-3.4641016151377553)-- (8.,3.4641016151377553);
\draw [line width=0.8pt,dotted] (6.,-3.4641016151377553)-- (9.,1.7320508075688776);
\draw [line width=0.8pt,dotted] (8.,-3.4641016151377553)-- (10.,0.);
\draw [line width=0.8pt,dotted] (10.,-3.4641016151377553)-- (11.,-1.7320508075688776);
\draw [line width=0.8pt,dotted] (2.,-3.4641016151377553)-- (0.,0.);
\draw [line width=0.8pt,dotted] (0.,-3.4641016151377553)-- (-1.,-1.7320508075688776);
\draw [line width=0.8pt,dotted] (1.,1.7320508075688776)-- (9.,1.7320508075688776);
\draw [line width=0.8pt,dotted] (2.,3.4641016151377553)-- (8.,3.4641016151377553);
\draw [line width=0.8pt,dotted] (3.,5.196152422706633)-- (7.,5.196152422706633);
\draw [line width=0.8pt,dotted] (4.,6.9282032302755105)-- (6.,6.9282032302755105);
\draw [line width=0.8pt,dotted] (4.,6.9282032302755105)-- (5.,5.19615242270663);
\draw [line width=0.8pt,dotted] (5.,5.19615242270663)-- (6.,6.9282032302755105);
\draw [line width=0.8pt,dotted] (1.,-1.7320508075688776)-- (0.,-3.4641016151377553);
\draw [line width=0.8pt,dotted] (1.,-1.7320508075688776)-- (-1.,-1.7320508075688776);
\draw [line width=0.8pt,dotted] (9.,-1.7320508075688776)-- (11.,-1.7320508075688776);
\draw [line width=0.8pt,dotted] (9.,-1.7320508075688776)-- (10.,-3.4641016151377553);
\draw [line width=0.8pt,dotted] (6.,0.)-- (10.,0.);
\draw [line width=0.8pt,dotted] (4.,0.)-- (0.,0.);
\draw [line width=0.8pt,dotted] (7.,5.196152422706633)-- (5.,1.7320508075688774);
\draw [line width=0.8pt,dotted] (4.,0.)-- (2.,-3.4641016151377553);
\draw [line width=0.8pt,dotted] (6.,0.)-- (8.,-3.4641016151377553);
\draw [line width=0.8pt,dotted] (3.,5.196152422706633)-- (5.,1.7320508075688774);
\draw (4.2,1.2) node[anchor=north west] {$T_0$};
\draw (4.2,3.6) node[anchor=north west] {$T_1$};
\draw (4.2,7.1) node[anchor=north west] {$T_2$};
\end{tikzpicture}
\caption{The partition of the plane into sectors  $T_n, n=0,1,2\ldots$ each containing $18n-3$ virtual domains.}
\label{sec2 : fig1a}
\end{figure}
The distance of each sector from the initial triangle is $(n-1)h$. Let $0<a<b$ then for every $n$ and $t\in (a,b)$ we have: 
\beqr
\int_{\triangle}\sum_{l=0}^{k} p_{0,t}(x,R'_l y)\,dx &\leq &\int_{\triangle}\sum_{l=0}^{\infty} p_{0,t}(x,R'_l y)\,dx \non \\&\leq &\frac{1}{4\pi \alpha} \int_{\triangle}\left(1+\sum_{n=1}^{\infty}(9n-1)\exp \left(-\frac{(n-1)^2 h^2}{4b}\right)\right)dx=C.
\label{sec2 : eq13c}
\feqr
Therefore from Lemma~\rf{sec2 : lem1} the function $p_t^{+\triangle}(x,y)$ is a smooth solution of the heat equation on $M\times (a,b)$. Since $(a,b)$ is arbitrarily chosen this property can be extended to the whole $(0,\infty)\times\Omega$ space. The second sum $p_t^{-\triangle}(x,y)$ can be treated similarly and therefore the claim is true.
\item $p_t^\triangle(\cdot,y)\xrightarrow{\mc{D'}(\triangle
)}\delta_y$ as $t\rightarrow 0$.

Let $f\in C_0^\infty(\Omega)$ then we have
\beqr
\lim_{t\rightarrow 0^+}\int_{\triangle} \left(\sum_{l=1}^{\infty} (-1)^{s(l)}p_{0,t}(x,R_l y)\right)f(x)\,dx \!=\!\lim_{k\rightarrow \infty}\sum_{l=1}^{k}\left(\lim_{t\rightarrow 0^+}\int_{\triangle} (-1)^{s(l)}p_{0,t}(x,R_l y)f(x)\,dx\right)\!=\!0, 
\label{sec2 : eq13d}
\feqr
where the first equality is justified by the uniform convergence of both $\sum_{l=1}^{\infty}(-1)^{s(l)}p_{0,t}(x,R_l y)f(x)$ on $x\in \triangle$ and $\sum_{l=1}^{k}\int_{\triangle} (-1)^{s(l)}p_{0,t}(x,R_l y)f(x)\,dx$ on $t\in (0,c)$, for $c$ small enough. The last equality is due to Lemma~\rf{sec2 : lem1a} since for every $l$ the point $R_l y\notin \overline{\triangle}$. The term $p_{0,t}(x,R_0 y)$ from Lemma~\rf{sec2 : lem1a} also leads to $p_{0,t}(\cdot,y)\xrightarrow{\mc{D'}(\triangle)}\delta_y$ as $t\rightarrow 0$ given that $y\in \triangle$. The first two statements assure that $p_t^\triangle(x,y)$ is the fundamental solution of the heat equation.
\item
$\limsup_{k\rightarrow\infty}p_{t_k}^\triangle(x_k,y)\geq 0$
for every sequence $(t_k,x_k)$ such that $t_k\rightarrow 0$ and  $x_k\rightarrow x\in \triangle$. This statement is the weaker condition satisfied by the fundamental solution in Theorem~\rf{sec1 : th2}.

For $x\in\triangle, t\in(0,c)$ and $c$ small, we have 
\beqr
\sum_{l=1}^{\infty}p_{0,t}(x,R_l y)\leq \sum_{k=1}^{\infty} \frac{(18k-3)}{4\pi c}\exp \left(-\frac{(k-1)^2 h^2}{4c}\right)<\infty.
\label{sec2 : eq13e}
\feqr
From the previous relation we conclude that the series $\sum_{l=1}^{\infty} (-1)^{s(l)}p_{0,t}(x,R_l y)$ uniformly converges for $x\in\triangle, t\in(0,c)$, therefore for every sequence $(t_k,x_k)$ such that $t_k\rightarrow 0$ and $x_k\rightarrow x\in \Omega$, we have
\beqr
\lim_{k\rightarrow \infty}\sum_{l=1}^{\infty} (-1)^{s(l)}p_{0,t_k}(x_k,R_l y)=\sum_{l=1}^{\infty} (-1)^{s(l)}\lim_{k\rightarrow \infty}p_{0,t_k}(x_k,R_l y)=\sum_{l=1}^\infty 0=0.
\label{sec2 : eq13f}
\feqr
Due to the inequality
$\limsup_{k\rightarrow\infty}p_{0,t_k}(x_k,y)\geq 0$
we end up with the desired result.
\item 
$p_t^{\triangle}(x,y)\rightrightarrows 0$ as $x\rightarrow \partial \triangle$ where the convergence is uniform w.r.t. $t\in (0,T)$ for every $T>0$.

The function $p_t^{\triangle}(x,y)$ vanishes on the boundary $\partial\triangle$ by construction. Let $x'$ be a point of the boundary $\partial\triangle$ such that $|x-x'|=d(x,\partial\triangle)$. Using the mean value theorem we obtain:
\beqr
\left|p_t^\triangle(x,y)\right|&=&\left|p_t^\triangle(x,y)-p_t^\triangle(x',y)\right| \non \\
&\leq& \left|\nabla p_{0,t}(z-y)\right| |x-x'|+ \sum_{l=1}^{\infty}\left|\nabla p_{0,t}(z_l-R_l y)\right| |x-x'|,
\label{sec2 : eq13g}
\feqr
where $z_l,z\in[x,x']$.
For $d(x,\partial\triangle)<d(y,\partial\triangle)/2$ we have
\beqr
|z-y|\geq d(y,\partial\triangle)-d(z,\partial\triangle)\geq d(y,\partial\triangle)-d(x,\partial\triangle)\geq \frac{d(y,\partial\triangle)}{2}.
\label{sec2 : eq13h}
\feqr
Also for those $l$'s satisfying $R_l y\in T_n$, it holds
\beqr
|z_l-R_l y|>(n-1)h+d(y,\partial\triangle).
\label{sec2 : eq13i}
\feqr
Using the previous relations and Lemma~\rf{sec2 : lem2} we get:
\beqr
&&\sup_{t>0}\left|p_t^\triangle(x,y)\right| \non \\
&&\leq \left(\sup_{\begin{subarray}{c}
t>0\\
|z-y|\geq d(y,\partial\triangle)/2
\end{subarray}}\left|\nabla p_{0,t}(z-y)\right|+\sum_{n=1}^{\infty}(18n-3)\sup_{\begin{subarray}{c}
t>0\\
|z'|>(n-1)h+d(y,\partial\triangle)
\end{subarray}}\left|\nabla p_{0,t}(z')\right|\right)|x-x'|\non \\
&&\leq\left(\frac{\textrm{const.}}{(d(y,\partial\triangle)/2)^3}+\sum_{n=1}^{\infty}(18n-3)\frac{\textrm{const.}}{((n-1)h+d(y,\partial\triangle))^3}\right)d(x,\partial\triangle) \non \\
&&=C d(x,\partial\triangle).
\label{sec2 : eq13k}
\feqr
Therefore by Theorem~\rf{sec1 : th2}, $u(t,x)\equiv p_t^{\triangle}(x,y)$. 
\end{enumerate}
\end{proof}
\par In the sequel we calculate the asymptotic behaviour of the partition function for the equilateral triangle and show how the reflection elements of $\textrm{Dih}_{6}$ group conspire to give the surface area and the length of its perimeter. Each vertex, its two adjacent median points and the centroid form three quadrilateral subdomains $D_i, \, i=1,2,3$. The Cartesian coordinates of the  triangle vertices are: $A=(0, 0)$, $B=(L, 0)$, $C=(L/2, \sqrt{3}L/2)$ and the reflection matrices are given by
\beqr
R_{a_1}(\frac{2\pi}{3})=\left( \begin{array}{cc} -\frac{1}{2} & \frac{\sqrt{3}}{2} \\ \frac{\sqrt{3}}{2} & \frac{1}{2} \end{array}\right)  \quad 
\textrm{and} \quad R_{a_2}=\left(\begin{array}{cc} 1 & 0 \\ 0 & -1 \end{array}\right) .
\label{sec2 : eq16}
\feqr 
The contribution of the identity element $I\equiv id$ to $Z_{\Omega}$ is
\beqr
\frac{1}{4\pi t}\sum_{i=1}^3\iint_{D_i} \!\!\! dx\,dy=\frac{1}{4\pi t} |\Omega|, 
\label{sec2 : eq17}
\feqr          
where $|\Omega|$ is the area of the equilateral triangle. The contribution of the reflection elements $R_{a_1},\, R_{a_2}$ and $R_{a_1}\cdot R_{a_2}\cdot R_{a_1}$ from the neighbouring to a vertex virtual domains (see Figure~\rf{chap2 : fig4}) is
\beqr
&&\!\!\!\!\!\!\!\!\!\!\lim_{t\rightarrow 0^+}\left(\sum_{i=1}^3\iint_{D_i}  p_t^{refl.}(x,y)dx\,dy\right)= 3\lim_{t\rightarrow 0^+}\left[\int_{0}^{\frac{L}{4}}\!\!\!\int_{0}^{x \sqrt{3}}\!\!\!p_t^{refl.}(x,y)dy \, dx+\int_{\frac{L}{4}}^{\frac{L}{2}}\!\!\!\int_{0}^{\frac{L-x}{\sqrt{3}}}\!\!\! p_t^{refl.}(x,y) dy \, dx\right] \non \\
&\stackrel{t\rightarrow 0^+}{\sim}&\!\!\!\!\!\!\!\! - \frac{|\partial \Omega|}{8\sqrt{\pi t}}, \quad \textrm{where} \quad
p_t^{refl.}(x,y)= -\frac{1}{4\pi t}\left(e^{-\frac{(\sqrt{3}x-y)^2}{4t}}+e^{-\frac{y^2}{t}}+e^{-\frac{(\sqrt{3}x+y)^2}{4t}}\right) 
\label{sec2 : eq18}
\feqr
and the integration formulas of Appendix A2 have been used. The rotations $R_{a_2}\cdot R_{a_1}$ and $(R_{a_2}\cdot R_{a_1})^2$ give
\beqr
\lim_{t\rightarrow 0^+}\left(\sum_{i=1}^3\iint_{D_i} \!\!\! p_t^{rot.}(x,y)dx\,dy\right) \!\!\!\!&=& \!\!\!\! 3\cdot 2 \lim_{t\rightarrow 0^+}\frac{1}{4\pi t}\biggl[\int_{0}^{\frac{L}{4}}e^{-\frac{3 x^2}{4t}} \left(\int_{0}^{x \sqrt{3}} e^{-\frac{3 y^2}{4t}}dy\right) \, dx \non \\
\!\!\!\! &+&\!\!\!\! \int_{\frac{L}{4}}^{\frac{L}{2}} e^{-\frac{3 x^2}{4t}} \left(\int_{0}^{\frac{L-x}{\sqrt{3}}} e^{-\frac{3y^2}{4t}} dy\right) \, dx\biggr]= \frac{1}{3}.
\label{sec2 : eq18a}
\feqr 
\begin{figure}[h]
\centering
\includegraphics[scale=0.45]{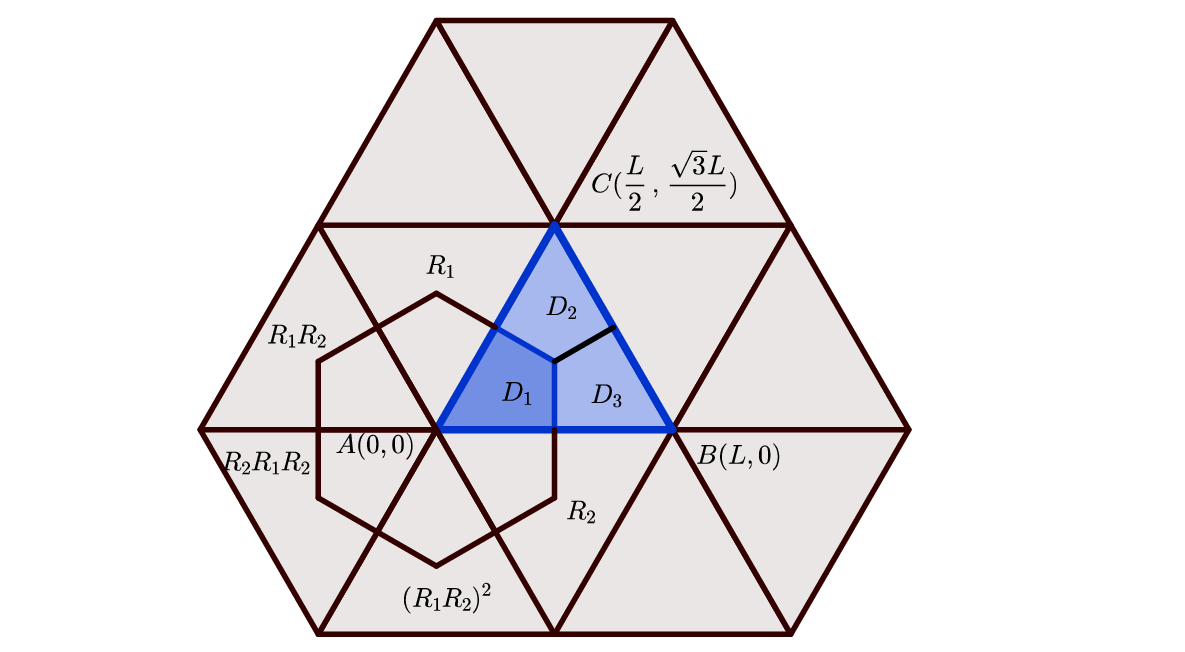} 
\caption{This figure depicts the virtual domains generated by the elements of the $\textrm{Dih}_{6}$ group for an equilateral triangle neglecting the next to neighbours regions.} 
\label{chap2 : fig4}
\end{figure}  

\textbf{Remarks}
\begin{enumerate}
\item Due to the defining relations \rf{sec2 : eq7} one can prove the following identities
\beqr
R_{a_1}\cdot R_{a_2}\cdot R_{a_1}=R_{a_2}\cdot R_{a_1}\cdot R_{a_2}, \quad (R_{a_2}\cdot R_{a_1})^2=R_{a_1}\cdot R_{a_2}. 
\label{sec2 : eq18b}
\feqr
Such relations support the simplification of the computations and depend on the anticlockwise (or clockwise) orientation of the reflections.
\item Applying \rf{sec2 : eq13} the topological terms for the triangles $(2,4,4)$, $(2,3,6)$ and for the rectangle are $3/8$, $5/12$ and $1/4$ respectively. Comparing the values of the topological term we establish the following decreasing sequence:
$c_{\textrm{rectangle}}<c_{(3,3,3)}<c_{(2,4,4)}<c_{(2,3,6)}$ for fixed area and perimeter. The previous inequalities indicate that the topological term is minimised by squares (or rectangles) while the equilateral triangles saturate the lowest bound among all acceptable triangles as one may prove by using the method of Lagrange undetermined multipliers.  
\end{enumerate}
\subsection{The \texorpdfstring{$d=3$}{d3} case}

In three dimensions the bounded regions compatible with the image method are: triangular prisms with $(\theta_{12}, \theta_{23}, \theta_{31})=(\pi/3, \pi/2, \pi/2)$,  right triangular prisms (both with five faces) and rectangular parallelipipeds (six faces). Again we study the infinite trihedral angle by applying the image method to determine the time independent term in $Z_{\Omega}$. Let $\theta_{ij}$ denote the angle between the i- and j-plane (see Figure~\rf{chap2 : fig5} ). 
\begin{figure}[ht]
\centering
\includegraphics[scale=0.7]{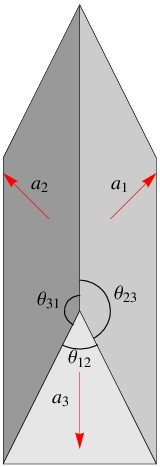} 
\caption{The angles $\theta_{ij}$ and the roots $a_i$ orthogonal to the $i$-plane.} 
\label{chap2 : fig5}
\end{figure}
Setting $\theta_{23}=\theta_{31}=\pi/2$, the reflection matrices associated with the roots $\alpha_i, \, i=1,2,3$ orthogonal to the i-planes are found to be
\beqr
R_{a_1}(\theta_{12})\!\!\!\!\!\!&=&\!\!\!\!\!\!\left(\begin{array}{ccc} \cos (2\theta_{12}) & \sin (2\theta_{12})  & 0 \\ 
\sin (2\theta_{12})  & -\cos (2\theta_{12}) & 0 \\  
0 & 0 & 1 \\ \end{array}
\right), \,\, R_{a_2}= \left(\begin{array}{ccc} 1 & 0 & 0\\
0 & -1 & 0 \\
0 & 0 & 1 \end{array}
\right), \,\,  R_{a_3}= \left(\begin{array}{ccc} 1 & 0 & 0\\
0 & 1 & 0 \\
0 & 0 & -1 \end{array}
\right) 
\label{sec2 : eq19} 
\feqr
where $\theta_{12}=\pi/p, \,p \in \mathbb{N}/ \{1\}$. The interior of the trihedral domain does not contain any virtual mirror generated by multiple reflections in the three bounding hyperplanes only if $(p,q,r)$ (or permutation of them) satisfy
\beqr
\frac{\pi}{p}+\frac{\pi}{q}+\frac{\pi}{r}>\pi \quad \textrm{and} \quad r, q, p \in \mathbb{N}\setminus \{1\}.
\label{sec2 : eq21a}
\feqr
The solutions to \rf{sec2 : eq21a} are 
\beqr
(2,2,r), \quad (2,3,3), \quad (2,3,4), \quad \textrm{and} \quad (2,3,5). 
\label{sec2 : eq22}
\feqr     
The defining relations of the corresponding group become
\beqr
\left(R_{a_1}\right)^2=\left(R_{a_2}\right)^2=\left(R_{a_3}\right)^2=\left(R_{a_1}\cdot R_{\alpha_2}\right)^p=\left(R_{a_2}\cdot R_{\alpha_3}\right)^q=\left(R_{a_3}\cdot R_{a_1}\right)^r=I.
\label{sec2 : eq23}
\feqr 
\begin{proposition}
\label{sec2 : prop3}
If the infinite trihedral angle is $(\theta_{12},\theta_{23},\theta_{31})=(\pi/r,\pi/2,\pi/2), \, r\in \mathbb{N}\setminus \{1\}$ then only the elements of $\mathcal{C}'_r \times \{ R_{a_3}\}\subset \textrm{Dih}_{2m}\times \mathbb{Z}/2\mathbb{Z}$ contribute to the constant term of $Z_{\Omega}$. In this case the contribution reads
\beqr
I_{const.}(r)=-\frac{1}{32r}\sum_{k=1}^{r-1} \frac{1}{\sin^2\left(\frac{k\pi}{r}\right)}=-\frac{1}{96 r}(r^2-1).
\label{sec2 : eq24}
\feqr  
\end{proposition}
\begin{proof}
Setting $\theta_{12}=\pi/r$ into relation \rf{sec2 : eq19}  we obtain
\beqr
R_{a_1}=\left(\begin{array}{ccc} \cos(\frac{2\pi}{r}) & \sin (\frac{2\pi}{r})  & 0 \\ 
\sin (\frac{2\pi}{r})  & -\cos(\frac{2\pi}{r})& 0\\  
0 & 0& 1 \\ \end{array}
\right).
\label{sec2 : eq26}
\feqr
The elements of $\mathcal{C}'_r \times \{ R_{a_3}\}$ have the matrix representation
\beqr
(R_{a_1}\cdot R_{\alpha_2})^{k}\cdot R_{a_3}=\left(\begin{array}{ccc} \cos(\frac{k\pi}{r}) & -\sin (\frac{k\pi}{r})  & 0 \\ 
\sin (\frac{k\pi}{r})  & \cos(\frac{k\pi}{r})& 0\\  
0 & 0& -1 \\ \end{array}
\right), \quad   k=1,\ldots, r-1
\label{sec2 : eq27}
\feqr
and are related to reflections. The heat kernel in this case is given by
\beqr
p_t(x,y,z)\sim -\left(\frac{1}{4\pi t}\sum_{k=1}^{r-1}e^{-\frac{1}{t}\sin^2\left(\frac{k\pi}{r}\right)(x^2+y^2)}\right)\left(\frac{1}{\sqrt{4\pi t}}e^{-\frac{z^2}{t}}\right). 
\label{sec2 : eq28}
\feqr 
The observed factorization is due to the direct product nature of the group $\mathcal{C}'_r \times \{ R_{a_3}\}$. Integration over the infinite trihedral angle produces \rf{sec2 : eq24}. 
\end{proof}
The rest of the group elements give the following transition densities: 
\begin{table}[h]
\centering
\begin{tabular}{|c|c||c|c|}\hline
\multicolumn{2}{|c||}{\bfseries Reflections} & \multicolumn{2}{|c|}{\bfseries Rotations} \\ \hline \hline
\itshape Group Element & \itshape $(4\pi t)^{\frac{3}{2}} p_t(x,y,z)$ & \itshape Group Element & $(4\pi t)^{\frac{3}{2}} p_t(x,y,z)$  \\ \hline 
$R_{a_1}$ & $-e^{-\frac{1}{t}(\sin (\frac{\pi}{r}) x - \cos(\frac{\pi}{r})y)^2}$ & $(R_{a_1}\cdot R_{a_2})^{k}$, $k=1,\cdots,r-1$  & $e^{-\frac{1}{t} \sin^2 (\frac{k\pi}{r})(x^2+y^2)}$ \\ \hline
$R_{a_2}$ & $-e^{-\frac{y^2}{t}}$ & $R_{a_3}\cdot R_{\alpha_1}$ & $e^{-\frac{1}{t}\left[(\sin (\frac{\pi}{r}) x - \cos(\frac{\pi}{r})y)^2+z^2\right]}$ \\ \hline
$R_{a_3}$ & $-e^{-\frac{z^2}{t}}$ & $R_{a_3}\cdot R_{\alpha_2}$ & $e^{-\frac{(y^2+z^2)}{t}}$ \\ \hline
\end{tabular}
\caption{The heat kernels of group elements which do not belong to $\mathcal{C}'_r \times \{ R_{a_3}\}$.}
\end{table}
\par According to Coxeter's classification scheme there are polyhedra which fall into the class of three-dimensional tessellations but the method of images puts severe constraints on the admissible ones through the requirement $\theta_{ij}=\pi/p_{ij}, \, p_{ij} \in \mathbb{N}\setminus \{1\},\, \forall i,j$. An example is the three tetrahedra $(0123),\,(0023)$ and $(0033)$ (see Figure~\rf{chap2 : fig6}). In particular for the $(0123)$ tetrahedron we have the angle $1\hat{0}_{\textrm{left}}3_{\textrm{up}}\backsimeq54.73^{\circ}$.  
\begin{figure}[h]
\centering
\psfrag{0}[c]{$0$}
\psfrag{1}[c]{$1$}
\psfrag{2}[c]{$2$}
\psfrag{3}[c]{$3$}
\includegraphics[scale=0.4]{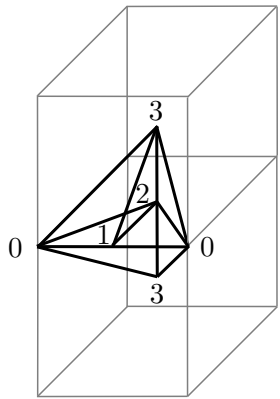} 
\caption{The symbol $(ijkl)$ denotes the tetrahedron with edges $(ij)$, $(jk)$, $(kl)$. The coordinates of the points $\{0,1,2,3\}$ are: $0_{\textrm{left}}=(0,0,0)$, $0_{\textrm{right}}=(L,0,0)$, $3_{\textrm{up}}=(\frac{L}{2},\frac{L}{2},\frac{L}{2})$, $3_{\textrm{down}}=(\frac{L}{2},\frac{L}{2},-\frac{L}{2})$, $1=(\frac{L}{2},0,0)$, $2=(\frac{L}{2},\frac{L}{2},0)$. The length of each side is given by $\overline{(ij)}=\frac{\overline{(00)}}{2}\sqrt{j-i}, \, j>i$.} 
\label{chap2 : fig6}
\end{figure}
\par The first candidate in our investigation is the prism $(\theta_{12}, \theta_{23}, \theta_{31})=(\pi/3, \pi/2, \pi/2)$, which in the light of Proposition~\rf{sec2 : prop3} gives the topological term $-1/6$. The asymptotic behaviour of the partition function is found to be 
\beqr
Z_{\Omega,(2,2,3)}(t)&\stackrel{t\rightarrow 0^+}{\sim}& \frac{\sqrt{3}a^2c}{4(4\pi t)^{\frac{3}{2}}}-\frac{a(a\sqrt{3}/2+3c)}{16\pi t}+\frac{3(2a+c)}{32\sqrt{\pi t}}-\frac{1}{6} 
\label{sec2 : eq30}
\feqr
where $a$ is the side length of the equilateral triangle and $c$ the height of the prism. 
\par The next candidates are the right triangular prisms with ($\theta_{12}=\theta_{23}=\theta_{31}=\pi/2$) at two vertices. We distinguish the following two possibilities 
\beqr
Z_{\Omega,(2,4,4)}(t)&\stackrel{t\rightarrow 0^+}{\sim}& \frac{a^2c}{2(4\pi t)^{\frac{3}{2}}}-\frac{a(a+c(2+\sqrt{2}))}{16\pi t}+\frac{2a(2+\sqrt{2})+3c}{32\sqrt{\pi t}}-\frac{3}{16} 
\label{sec2 : tp1} \\
Z_{\Omega,(2,3,6)}(t)&\stackrel{t\rightarrow 0^+}{\sim}& \frac{a^2c \sqrt{3}}{2(4\pi t)^{\frac{3}{2}}}-\frac{a(a\sqrt{3}+c(3+\sqrt{3}))}{16\pi t}+\frac{2a(3+\sqrt{3})+3c}{32\sqrt{\pi t}}-\frac{5}{24}
\label{sec2 : tp2}
\feqr
where $a$ in \rf{sec2 : tp2} is the length of the side opposite to the $\pi/6$ angle and $c$ as in the previous case.
\par Finally, we study the rectangular parrallelipiped with edge lengths $(a,b,c)$. Dividing it into eight sub-rectangular parrallelipipeds we arbitrarily choose one of them and perform our computation. Their contributions are listed in the following table:
\begin{table}[H]
\centering
\begin{tabular}{|c|c||c|c|}\hline
\multicolumn{2}{|c||}{\bfseries Reflections} & \multicolumn{2}{|c|}{\bfseries Rotations} \\ \hline \hline
\itshape Group Element & \itshape Contribution  & \itshape Group Element & \itshape Contribution  \\ \hline 
$R_{\alpha_1}$ & $-\frac{bc}{64 \pi t} \textrm{erf}\left(\frac{a}{2\sqrt{t}}\right)$ & $R_{\alpha_1}\cdot R_{\alpha_2}$  & $\frac{c}{64 \sqrt{\pi t}} \textrm{erf}\left(\frac{a}{2\sqrt{t}}\right)\textrm{erf}\left(\frac{b}{2\sqrt{t}}\right)$ \\ \hline
$R_{\alpha_2}$ & $-\frac{ac}{64 \pi t} \textrm{erf}\left(\frac{b}{2\sqrt{t}}\right)$ & $R_{\alpha_3}\cdot R_{\alpha_1}$ & $\frac{b}{64 \sqrt{\pi t}} \textrm{erf}\left(\frac{a}{2\sqrt{t}}\right)\textrm{erf}\left(\frac{c}{2\sqrt{t}}\right)$  \\ \hline
$R_{\alpha_3}$ & $-\frac{ab}{64 \pi t} \textrm{erf}\left(\frac{c}{2\sqrt{t}}\right)$ & $R_{\alpha_3}\cdot R_{\alpha_2}$ & $\frac{a}{64 \sqrt{\pi t}} \textrm{erf}\left(\frac{b}{2\sqrt{t}}\right)\textrm{erf}\left(\frac{c}{2\sqrt{t}}\right)$  \\ \hline
\end{tabular}
\caption{The heat kernels and the associated reflection and rotation group elements.}
\end{table}
The identity element again gives the volume of the domain, the reflection elements $R_{a_i},\, i=1,2,3$ provide the area of the bounding surfaces, the rotations contribute to the lengths of the edges and the reflection  $R_{a_1}\cdot R_{a_2}\cdot R_{a_3}$ supports the topological term. In the short-time limit the error functions converge to unity and if we sum up the contributions from all the vertices we recover the correct result
\beqr
Z_{\Omega}(t)\stackrel{t\rightarrow 0^+}{\sim} \frac{abc}{(4\pi t)^{\frac{3}{2}}}-\frac{2(ab+bc+ac)}{16\pi t}+\frac{4(a+b+c)}{32\sqrt{\pi t}}-\frac{1}{8}.
 \label{sec2 : rp2}
\feqr 
Again depending on the shape of the polytope we observe the following decreasing sequence of values for the topological term:
$c_{\textrm{rec. par.}}<c_{(2,2,3)}<c_{(2,4,4)}<c_{(2,3,6)}$ for fixed volume, boundary and coboundary.    
\section{Conclusions}
In this paper, using a group theoretic approach, we studied analytically the short-time asymptotics of the free partition function corresponding to the Dirichlet Laplacian on tessellations which possess mirror symmetry through the hyperplanes bounding the fundamental domain. Implementing the method of images along with the path integral representation of the heat kernel up to three-dimensions, we established the connection of the geometrical quantities $|\Omega|, |\partial \Omega|$ and the topological term, with certain elements of the orthogonal group. We proved rigorously that \rf{sec1 : eq13} is the actual heat kernel on $\Omega$. This method can be generalised to higher dimensions and may help to solve the same problem with Dirichlet fractional Laplacians without requiring the knowledge of its spectra.  

\section*{Acknowledgements}
We would like to thank O. Corradini for clarifying a crucial issue related to the worldline approach.

\addcontentsline{toc}{subsection}{Appendix A }
\section*{Appendix A }
\label{apA}
\renewcommand{\theequation}{A.\arabic{equation}}
\setcounter{equation}{0}
\subsection*{A1.}
\label{A1}
\textbf{Proof of \rf{sec1 : eq10a}}\\
For a polygon $P=\nu_1 \nu_2 \cdots \nu_m$ in $\mathbb{E}^2$ the curvature of $P$ at each interior vertex $\nu_i$ is the real number
\beqr
k_{\nu_i}=<\overrightarrow{\nu_{i-1}\nu_i},\overrightarrow{\nu_{i}\nu_{i+1}})=\theta_i
\label{ap0}
\feqr 
which represents the angle between the unit vectors $\overrightarrow{\nu_{i-1}\nu_i}$ and $\overrightarrow{\nu_{i}\nu_{i+1}}$. The global (or total curvature) of a polygon is defined to be the sum of angles of its consecutive edges \cite{Mor}
\beqr
\mathcal{K}(P)=\sum_{i=1}^{m-2}<\overrightarrow{\nu_{i}\nu_{i+1}},\overrightarrow{\nu_{i+1}\nu_{i+2}}).
\label{ap01}
\feqr  
Using the well-known result \cite{F} that $\mathcal{K}(P)=2\pi$ for a polygonal planar, convex and closed curve in $\mathbb{E}^2$, it is trivial to prove \rf{sec1 : eq10a}.
\subsection*{A2.}
In the computations we have made extensive use of the following integral representation of the error function and integrals involving the error function, exponentials and powers \cite{NG}
\beqr
\textrm{erf}(az)  &=& \frac{2az}{\sqrt{\pi}} \int_0^1 e^{-a^2z^2u^2} du, \,\, z=\mathcal{R}e \, z+i\mathcal{I}m \, z=x+iy\label{a1a} \\
\int_0^{\infty} \textrm{erf}(ax) e^{-b^2 x^2} \, dx &=& \frac{\sqrt{\pi}}{2b}-\frac{1}{b\sqrt{\pi}} \tan^{-1}\left(\frac{b}{a}\right) \label{a2aa} \\
\int_0^{\infty} x\,  \textrm{erf}(ax) e^{-b^2 x^2} \, dx &=& \frac{a}{2b^2} \frac{1}{\sqrt{a^2+b^2}}, \quad \mathcal{R}e(b^2)>\mathcal{R}e(a^2), \,\, \mathcal{R}e(b^2)>0. \label{a2ab}
\feqr
\section*{Appendix B }
\label{apB}
\renewcommand{\theequation}{B.\arabic{equation}}
\setcounter{equation}{0}
\textbf{Proof of} \textbf{$\sum_{k=1}^{m-1}\frac{1}{\sin^2\left(\frac{k \pi }{m}\right)}=\frac{1}{3}(m^2-1)$}\\

The finite sum is written equivalently as
\beqr
\sum_{k=1}^{m-1}\frac{1}{\sin^2\left(\frac{k \pi }{m}\right)}=m-1+\sum_{k=1}^{m-1}\cot^2\left(\frac{k \pi }{m}\right)
\label{a4}
\feqr
therefore it is enough to prove that the second sum equals $(m-1)(m-2)/3$. Using simple trigonometric identities we obtain
\beqr
\sum_{k=1}^{m-1}\cot^2\left(\frac{k \pi }{m}\right)=-\sum_{k=1}^{m-1}\left(\frac{1+\omega^k}{1-\omega^k}\right)^2, \quad \omega=e^{\frac{2i\pi}{m}}.
\label{a5}
\feqr 
Defining the function
\beqr
((x)):=\left\{\begin{array}{cc} x-\lfloor x \rfloor-\frac{1}{2}, & x\notin \mathbb{Z} \\ 0, & \textrm{otherwise} \end{array}\right.,
\label{a6}
\feqr
the function $((l/m))$ has period $m$ and therefore it can be expanded in finite Fourier series as
\beqr
\left(\left( \frac{l}{m}\right)\right)&=&\sum_{k=0}^{m-1}\hat{f}(k) \omega^{kl}, \quad \textrm{where} \non \\
\hat{f}(k)&=&\frac{1}{m}\sum_{l=0}^{m-1} \left(\frac{l}{m}-\frac{1}{2}\right)\omega^{-kl}=\left\{\begin{array}{cc} 0, & k=0 \\ \frac{1}{2m}\left(\frac{1+\omega^k}{1-\omega^k}\right), & \textrm{for k, m coprime and}\, k,m \in \mathbb{Z}_+.  \end{array}\right.
\label{a7}
\feqr
Applying the convolution theorem for finite Fourier series
\beqr
(f*g)(l)=\sum_{n=0}^{m-1}f(l-n)g(n) =m\sum_{n=0}^{m-1} \hat{f}(n)\hat{g}(n)\omega^{ln}
\label{a9}
\feqr 
for $l=0$ and $f=g$ we get
\beqr
(f*f)(0)=-\sum_{n=0}^{m-1}f^2(n) =m\sum_{n=0}^{m-1} \hat{f}^2(n)
\label{a10}
\feqr 
Thus \rf{a4} becomes
\beqr
\sum_{k=1}^{m-1}\cot^2\left(\frac{k \pi }{m}\right)=4m\sum_{n=1}^{m-1}\left(\frac{n}{m}-\frac{1}{2}\right)^2=\frac{1}{3}(m-1)(m-2).
\label{a11}
\feqr
\bibliographystyle{plain}

\end{document}